\documentclass[a4paper,fleqn,usenatbib]{mnras}

\usepackage{mathptmx}

\usepackage[T1]{fontenc}
\usepackage{ae,aecompl}

\usepackage{graphicx}	% Including figure files
\usepackage{amsmath}	% Advanced maths commands
\usepackage{amssymb}	% Extra maths symbols
\usepackage{bm}
\usepackage{color}
\usepackage[all]{hypcap}
\usepackage{hyperref}
\usepackage{multicol}
\usepackage{multirow}
\usepackage{flushend}
\usepackage{txfonts}
\usepackage[capitalize,nameinlink]{cleveref}

\usepackage{epstopdf}
\usepackage{float}
\usepackage{subcaption}

\title[X-ray variability characteristics  of  RE~J1034+396]{Long term X-ray variability characteristics of the narrow-line Seyfert 1 galaxy RE~J1034+396}

\author[K. Chaudhury et al.]{K. Chaudhury,$^{1,8}$ \thanks{E-mail:chaudhurykishor@gmail.com}
V. R. Chitnis,$^{2}$
A. R. Rao,$^{3}$
K. P. Singh,$^{4}$
Sudip  Bhattacharyya,$^{3}$
\newauthor
G. C. Dewangan,$^{5}$
S. Chakraborty,$^{3}$
S. Chandra,$^{7}$
G. C. Stewart,$^{6}$
K. Mukerjee,$^{3}$
\newauthor
and R. K. Dey$^{8}$
\\
\\
$^{1}$Department of Physics, Alipurduar College, Alipurduar, West Bengal, 736122, India\\
$^{2}$Department of High Energy Physics, Tata Institute of Fundamental Research, 1 Homi Bhabha Road, Mumbai 400005, India\\
$^{3}$Department of Astronomy and Astrophysics, Tata Institute of Fundamental Research, 1 Homi Bhabha Road, Mumbai 400005, India\\
$^{4}$Indian Institute of Science Education and Research Mohali, Knowledge city, Sector 81, SAS Nagar, Manauli, India\\
$^{5}$Inter-University Centre for Astronomy and Astrophysics, Ganeshkhind, Pune 411 007, India \\
$^{6}$Department of Physics and Astronomy, The University of Leicester, University Road, Leicester, LE1 7RH, United Kingdom\\
$^{7}$Centre for Space Research, North-West University, Potchefstroom, 2520, South Africa\\
$^{8}$Department of Physics, University of North Bengal, Siliguri, WB, 734013, India%\\
}

\date{Accepted  .... Received ...; in original form ...}

\pubyear{2015}

\begin{document}
\label{firstpage}
\pagerange{\pageref{firstpage}--\pageref{lastpage}}
\maketitle

% Abstract of the paper
\begin{abstract}
We present the results of our study of the long term X-ray variability characteristics of the
Narrow Line Seyfert 1 galaxy RE J1034+396. We use data obtained from the AstroSat satellite
along with the light curves obtained from XMM-Newton and Swift-XRT. We use the 0.3 - 7.0 keV 
and 3 - 20 keV data, respectively, from the SXT and the LAXPC of AstroSat. The
X-ray spectra in the 0.3 - 20 keV region are well fit with a model consisting of a power-law and
a soft excess described by a thermal-Compton emission with a large optical depth, consistent
with the earlier reported results. We have examined the X-ray light curves in the soft and
hard X-ray bands of SXT and LAXPC, respectively, and find that the variability is slightly
larger in the hard band. To investigate the variability characteristics of this source at different
time scales, we have used X-ray light curves obtained from XMM-Newton data (200 s to 100
ks range) and Swift-XRT data (1 day to 100 day range) and find that there are evidences to
suggest that the variability sharply increases at longer time scales. We argue that the mass
of the black hole in RE J1034+396 is likely to be $\sim$3 $\times$ 10$^6$ M$_\odot$, based on the similarity of
the observed QPO to the high frequency QPO seen in the Galactic black hole
binary, GRS 1915+105.
\end{abstract}

% Select between one and six entries from the list of approved keywords.
% Don't make up new ones.

%%%%%%%%%%%%%%%%%%%%%%%%%%%%%%%%%%%%%%%%%%%%%%%%%%

%%%%%%%%%%%%%%%%% BODY OF PAPER %%%%%%%%%%%%%%%%%%

\section{Introduction}

The active galactic nucleus (AGN) RE~J1034+396 (also known as Zw212.025) at redshift z=0.042 is a narrow line Seyfert 1-galaxy (NLS1) \citep{Pounds95, Mason96, Puchnar98, Breeveld98}. It was first observed during   the  ROSAT WFC all sky survey in 1990 \citep{Pounds93}. It has since been observed on many occasions at different wavelengths from radio to X-rays \citep{Puchnar95, Puchnar01, Pounds95, Breeveld98}. Further observations include in X-rays by ASCA (GIS \& SIS) in 1994 \citep{Serlem95}, the Deep Survey Spectrometer on board EUVE in 1997, the BeppoSAX narrow field instrument in 1997 \citep{Puchnar01}, by XMM-Newton in 2007 \citep{Middle09} and in the optical with the twin armed ISIS spectrograph on the William Herschel Telescope in La Palma.

The 0.1-2.4 keV spectrum observed with the ROSAT Position Sensitive Proportional Counter \citep[][PSPC]{Pfeffer86}, showed an unusual "big blue bump (BBB)"   with  a high temperature whose high energy turnover is observed at around 0.4 keV in the soft X-ray band \citep{Puchnar95, Puchnar98}. The BBB has a very high temperature ($kT~ \sim$ 100 eV), dominating the EUV and soft X-ray emission and  leaving a bare, power-law like continuum component in the optical \citep{Puchnar95, Puchnar97}. The lack of BBB emission in the UV as suggested by IUE data imply a very hot accretion disc around a relatively low mass black hole ($\sim 10^6$ solar masses, $M_\odot$) \citep{Puchnar95}. The strong and variable ultra-soft X-ray excesses could be explained by a high mass accretion rate onto a relatively low-mass black hole \citep{Pounds95}. Examining a wide band Spectral Energy Distribution (SED) of this source \citet{Done12} suggest that the spectrum consists of three components: a black body from the disk (representing the BBB), a hard coronal component (power law at high energies) and a low temperature high optical depth Comptonization of the disc emission in the soft X-ray region.  

The data from a  long XMM-Newton observation (91 ks) showed  a significant QPO (quasi-periodic oscillations) signal ($\nu$ = 2.7 $\times 10^{-4}$ Hz, corresponding to a period of about 1h) in the X-ray power spectrum for RE~J1034+396 \citep{Gierli08}, similar to the QPOs seen in the X-ray power spectrum of Black Hole Binaries  (BHBs) \citep{Middle09, Bian10}. As active galactic nuclei (AGNs) and quasars are thought to be scaled-up versions of Galactic black hole binaries, powered by accretion onto super-massive black hole with masses of $10^6-10^9$$M_\odot$  \citep{Gierli08}, the QPOs seen in this source signifies that accretion properties of  AGN are  similar to the  accretion flow around the BHBs \citep{Middle09}.

In this paper we present the results from our investigation of the long term X-ray variability characteristics of  RE~J1034+396 using  AstroSat data along with archival XMM-Newton and Swift-XRT data.  
%with Soft X-ray Telescope (SXT) in soft X-ray band  (0.3 - 7 keV) and Large Area X-ray Proportional Counters (LAXPC)
% in hard X-ray band  (3 - 20 keV) to understand the geometry of  the X-ray continuum emission and study the underlying physics. 
In the next section we present the AstroSat observations   and in \S 3, we present the variability study using  archival data from XMM-Newton and Swift-XRT. In the last section we discuss the results in the context of the mass of the black hole in this source.

\section{AstroSat Observations}

AstroSat, the first Indian multi wavelength space observatory successfully launched on September 28, 2015,
has five scientific instruments on board \citep{Singh14}. These are a Soft X-ray focusing Telescope (SXT), a
Large Area X-ray Proportional Counters (LAXPC), a Cadmium Zinc Telluride Imager (CZTI), a Scanning Sky Monitor 
(SSM) and an UltraViolet Imaging Telescope (UVIT). RE~J1034+396 was observed by AstroSat on 2016 April 21-22 (Observation ID : G05\_238T03\_9000000424).
In the present work data from the SXT and LAXPC are used.

%The primary instruments are the Soft X-ray focusing Telescope (SXT) which covers the soft X-ray and the Large Area X-ray Proportional Counters (LAXPC) which covers the hard X-ray wavebands. RE~J1034+396 was also detected by SXT \& LAXPC on board AstroSat on 2016 April 21-22. In this work we have used the data from SXT \& LAXPC on board AstroSat. 

\subsection{SXT Observations}

The soft X-ray Telescope (SXT) on board AstroSat is a grazing incidence X-ray telescope with focal 
length of 2 m with a thermoelectrically cooled CCD in the focal plane.  
%The detector used in the focal plane of the telescope is the Charged Coupled Device, CCD-22 of e2V Technologies Inc., UK. 
The effective area of the SXT is $\sim 65~cm^{2}$ at 1.5 keV. It covers the energy range of 0.3-8 keV, with 
an energy resolution of 5-6\% at 1.5 keV. It has a field of view of $\sim 40^{'}$ diameter and the angular
resolution is $2^{'}$ FWHM. The point spread function is described by a double King function with FWHM of 
40$^{''}$ for the inner core \citep{kp16, kp17}.  RE~J1034+396 was observed with the SXT in  
Photon Counting (PC) mode.  

% The level 1 data from individual orbits are received at the SXT POC (Payload
%Operation Centre) from the ISSDC (Indian Space Science Data Center). These are then 
%processed using the SXTPIPELINE at the POC which implements screening of events by RAM angle (the angle between the payload axis to the
%velocity vector direction of the spacecraft), contamination by charged particles from excursions through 
%the SAA and the effects of bright Earth. Further, all events with grades > 12 are removed, and 
%level2 data for each orbit produced. These level 2 data also include quick look products such as an image, 
%spectrum and light curve for the entire field of view. Using a python script developed by the
%SXT team, good time intervals (GTI's) during  each orbit were selected, time overlaps between consecutive
%orbit data files were  removed and a merged event file of all cleaned events was generated. This resulted in an exposure of $\sim 42.8$ks.

{\bf SXT observations are carried out with Sun avoidance angle $\ge$ 45$^\circ$ 
and RAM angle (the angle between the payload axis to the velocity vector 
direction of the spacecraft) > 12$^\circ$ to ensure the safety of the 
instrument. The level 1 data from individual orbits are received at the 
SXT POC (Payload Operation Centre) from the ISSDC (Indian Space Science 
Data Center). These are then processed using the SXTPIPELINE at the POC. 
The SXTPIPELINE does the event extraction, the time tagging of events, the coordinate 
transformation from raw (detector) to sky coordinates, bias subtraction and adjustment, 
flagging of bad pixels and calibration source events, event grading (grade 
definition similar to Swift-XRT, see Romano et al. 2005),  the PHA construction 
for each event, the conversion from the event PHA to PI and a search for hot and flickering pixels. Further screening criteria such as selecting events with 
bright Earth avoidance angle of $\ge$ 110 degrees, removing data taken during the 
passage through the South Atlantic Anomaly (SAA) using the condition that the Charged 
Particle Monitor (CPM) rate below 12 counts s$^{-1}$ are applied. Events with 
grades > 12 are also removed. Good Time Interval (GTI) files and level 2 quick 
look products such as an image, spectrum and light curve are produced 
for the entire field of view for each orbit \citep{kp16, kp17}. Using 
a python script developed by the SXT team, good time intervals (GTI's) during
each orbit were selected, time overlaps between consecutive orbit data
files were removed and a merged event file of all cleaned events was
generated. This resulted in an exposure of  42.8 ks.}

\begin{figure}
	% To include a figure from a file named example.*
	% Allowable file formats are eps or ps if compiling using latex
	% or pdf, png, jpg if compiling using pdflatex
	\centerline{\includegraphics[width=0.8\linewidth]{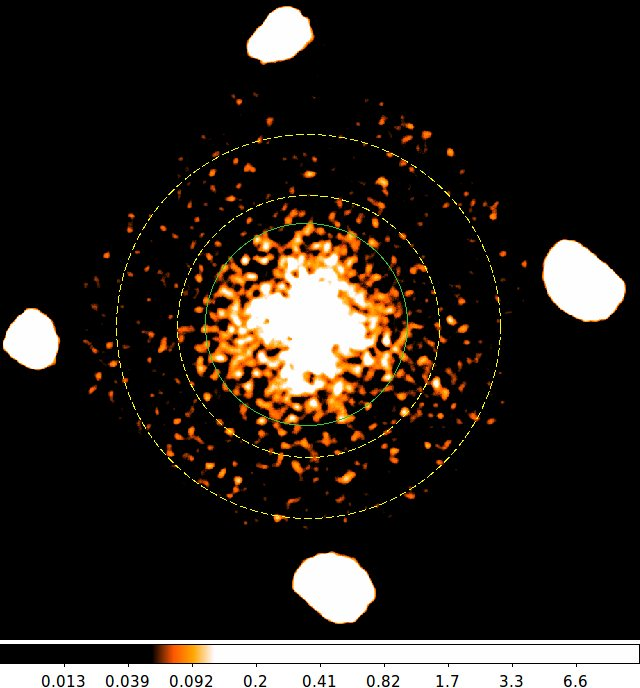}}
    \caption{AstroSat SXT image of RE~J1034+396 in the 0.3-8 keV energy band. The calibration sources 
located at the four corners are seen. Innermost circle shows the source region of radius 10$^{'}$ 
and the dashed annular region represents the background region with inner radius of 13$^{'}$ and outer 
radius of 19$^{'}$.}
    \label{fig:s}
\end{figure}

\begin{figure}
	% To include a figure from a file named example.*
	% Allowable file formats are eps or ps if compiling using latex
	% or pdf, png, jpg if compiling using pdflatex
	\centerline{\includegraphics[width=.68 \linewidth, angle=270 ]{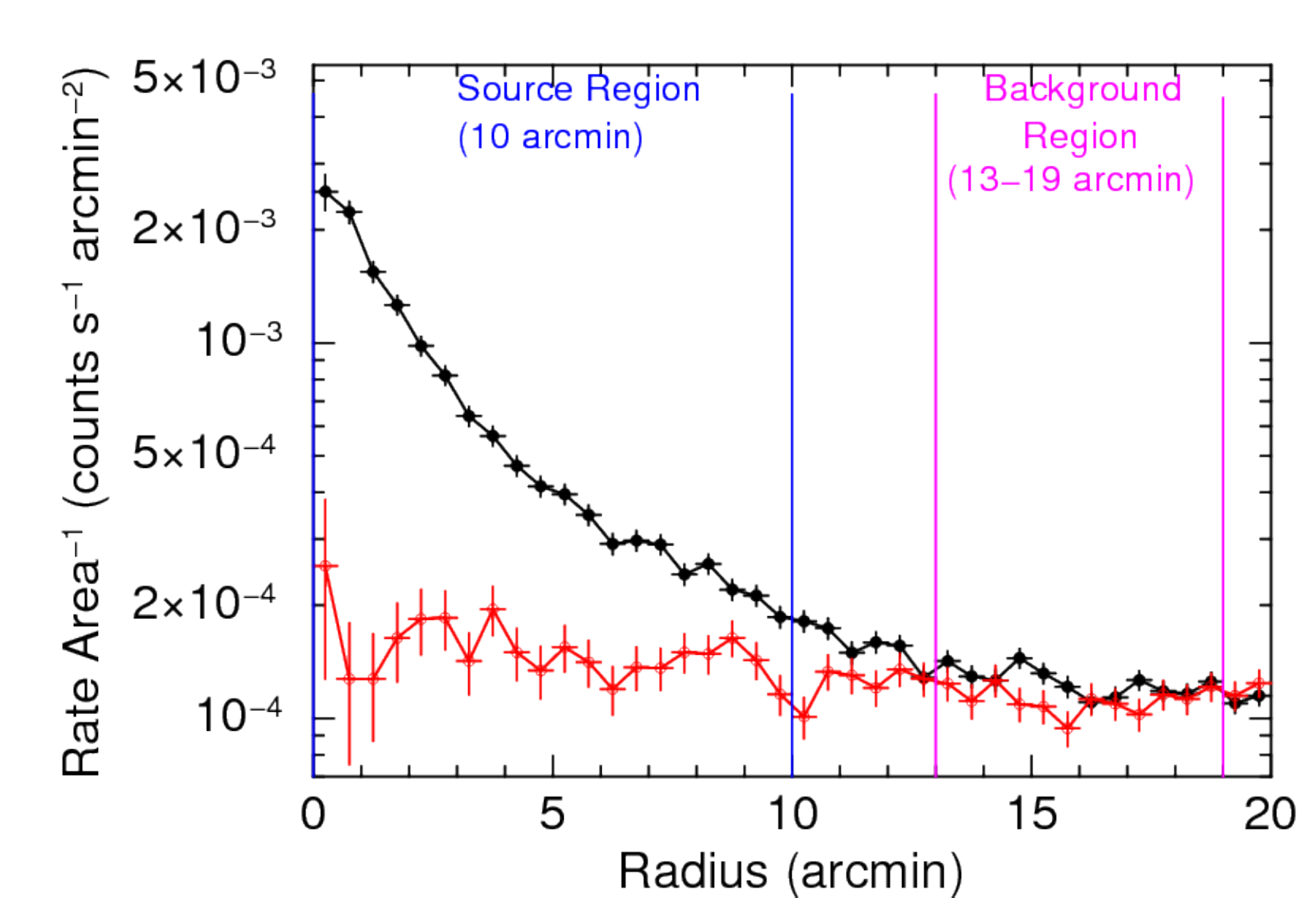}}
    \caption{The annular count distribution from the source using AstroSat SXT is shown with black dots connected with a line and the source free sky region  with red dots connected by a line. {\bf Error bars correspond to 1 $\sigma$.}}
    \label{fig:radial}
\end{figure}

Using the cleaned event
file an image was created (\Cref{fig:s}), and source and background regions were chosen. 
For this, XSELECT was used to extract light curves for 40 circles centred on the source with radii increasing in  steps of 0.5'.
Average count rates in the  0.3-7 keV  energy range were estimated and
surface brightness (counts per sec per unit sky area) values were calculated for the
40 annular regions.
%For this purpose, 
%circles of radii  0.5', 1.0', 
%1.5', 2.0', ........, 20.0' were considered and using XSELECT tool light curves were generated for
%each of these circles and average count rates in the energy range of 0.3-8 keV were estimated. Then
%photon flux (counts per sec per unit area) values were calculated for annular regions with radii 0'-0.5', 
%0.5'-1.0', 1.0'-1.5', ...., 19.5'-20.0'. 
The variation of these photon fluxes with radius is shown in \Cref{fig:radial}. The variation of the photon flux 
with radius obtained from a source free sky
image analysed in the same way is also shown. {\bf Error bars shown in the figure correspond to 1 $\sigma$. The total
count rate in the annular region with inner and outer radii of 13' and 19' is estimated to be (1.26$\pm$0.10)$\times$10$^{-4}$
counts s$^{-1}$ arcmin$^{-2}$ and (1.12$\pm$0.09)$\times$10$^{-4}$ counts s$^{-1}$ arcmin$^{-2}$ respectively 
for the source
and background runs. Beyond 10' source and background count rates match within errors.} 
Hence we have used a region with a radius of 10' around
RE~J1034+396 location as the source region and an annular region with an inner radius of 13' and an outer radius
of 19' as the background.

%We extracted data of RE~J1034+396, observed by SXT on board AstroSat of observation Id. G05\_238T03\_9000000424 from a tar file. Later data was merged to produce a event file using python script which is developed by SXT group. Then the XSELECT tool of HeaSoft package is used to create image file from the merged event file. Before extracting the X-ray light curve and spectrum, we estimated the source and background region in image file using ds9 (an imaging tool in HEASOFT package) \Cref{fig:s}. For proper source and background estimation we considered circles of radius 0.5, 1.0, 1.5, 2.0, ........20.0 arcmin and then calculated the count rate for each circle in energy range .3-8 keV using XSELECT tool. We calculated the photon flux (number of photons per second per unit area) using the following formula

%\begin{equation}
% \textrm{Photon Flux = }\frac{\textrm{count rate}}{\left({r_\mathrm{2}}^2 - {r_\mathrm{1}}^2\right)},
%\end{equation}      

%\begin{equation}
%\textrm{Mid Radius (R) = }\frac{\left(r_\mathrm{1} \textrm{+} r_\mathrm{2}\right)}{2},
%\end{equation}

The SXT source and background light curves over the energy range of 0.3-7 keV with a bin size of 5 seconds were 
extracted for each orbit using  these regions. A background subtracted
source light curve was generated for each orbit and the average source count rate for each  orbit estimated. The orbit-wise
light curve with a typical exposure of $\sim 2$ ks per bin is shown in the upper panel of \Cref{fig:olcsl20}.

To quantify the variability, the fractional variability amplitude $F_{var}$ given by

\begin{equation}
F_{var}=\sqrt{\frac{S^2-\overline{\sigma_i^2}}{\overline{x}^2}}
\label{Fvar}
\end{equation}

where $S^2$ is sample variance, $\overline{x}$ is mean rate and $\overline{\sigma_i^2}$ is
average variance from measurements for data with N samples was calculated \citep{chitnis09, vaughan03}.

The error on $F_{var}$ is then

\begin{equation}
err(F_{var}) = \frac{1}{\sqrt{2N}}\frac{S^2}{\overline{x}^2 F_{var}}
\end{equation}

From our SXT light curve, we estimate $F_{var}$ to be 0.123$\pm$0.031.

\subsection{LAXPC Observations} 

The AstroSat LAXPC consists of three identical proportional counter units covering the energy 
range of 3-80 keV. The energy resolution is $\sim$ 10-12\% in the 22-60 keV range, the field of view is
$0.9\degr\times 0.9\degr$ and the total effective area is $\sim$ 6000 cm$^2$ at 5-20 keV \citep{Singh14, Yadav16, 
Agrawal17, Antia2017}. The misalignment between the SXT and LAXPC instruments is 
$\sim$6' {\bf(for details see document on 'Relative alignment of AstroSat payloads'\footnote{\url{http://astrosat-ssc.iucaa.in/?q=documents}})} 
much smaller than the FOV of either.

The LAXPC  data were processed at the POC using code developed by the LAXPC team. This code 
combines the data from multiple orbits, removing overlapping data between consecutive orbit data files.
It generates an event file containing all events detected in addition to a light curve, 
spectrum and a list of GTI intervals. These intervals are defined as periods 
when the source elevation is $> 3^\circ.5$  above the Earth's limb, exclude
the SAA region, which is defined to be  times with the satellite longitude greater than 251$^\circ$.
Suitable  background files, based on a model built from the observed source free sky regions, are also generated.  
Data from source free sky regions observed within a few days of the source observation are used and
appropriate scaling is performed depending on the orbit. A fit to the variation of background 
counts as a function of latitude and longitude is used. In the case of a difference in the gain between the
source and background spectral files, an appropriate gain shift is applied to the background file.

To generate the light curve, source and background spectra are produced for each orbit.
To improve the statistics, data from only the top layers are used. 
Data from the three LAXPC units (called LX10, LX20 and LX30) are analysed separately. LX30 was {\bf suspected 
to have undergone
a gas leakage} resulting in a continuous gain shift {\bf \citep{Antia2017}} and hence these data were not used in the present  analysis.
Orbit-wise source and background
spectra are compared in the 35-80 keV region. In the top layer, no source counts are expected in this
energy range. For LAXPC unit LX20, the ratios of source to background counts in various pulse height (pha) channels were found
to be consistent with a constant. Hence the ratio of total counts from source to background spectra
was used as the normalisation factor for each orbit. It was found to vary from 0.94 to 1.02.
For LX10, the channel wise ratio showed a linear trend making the procedure for correction
difficult. We therefore have not used data from LX10 for  further analysis.

\begin{figure}
	% To include a figure from a file named example.*
	% Allowable file formats are eps or ps if compiling using latex
	% or pdf, png, jpg if compiling using pdflatex
	\centerline{\includegraphics[width=0.62 \linewidth, angle=270]{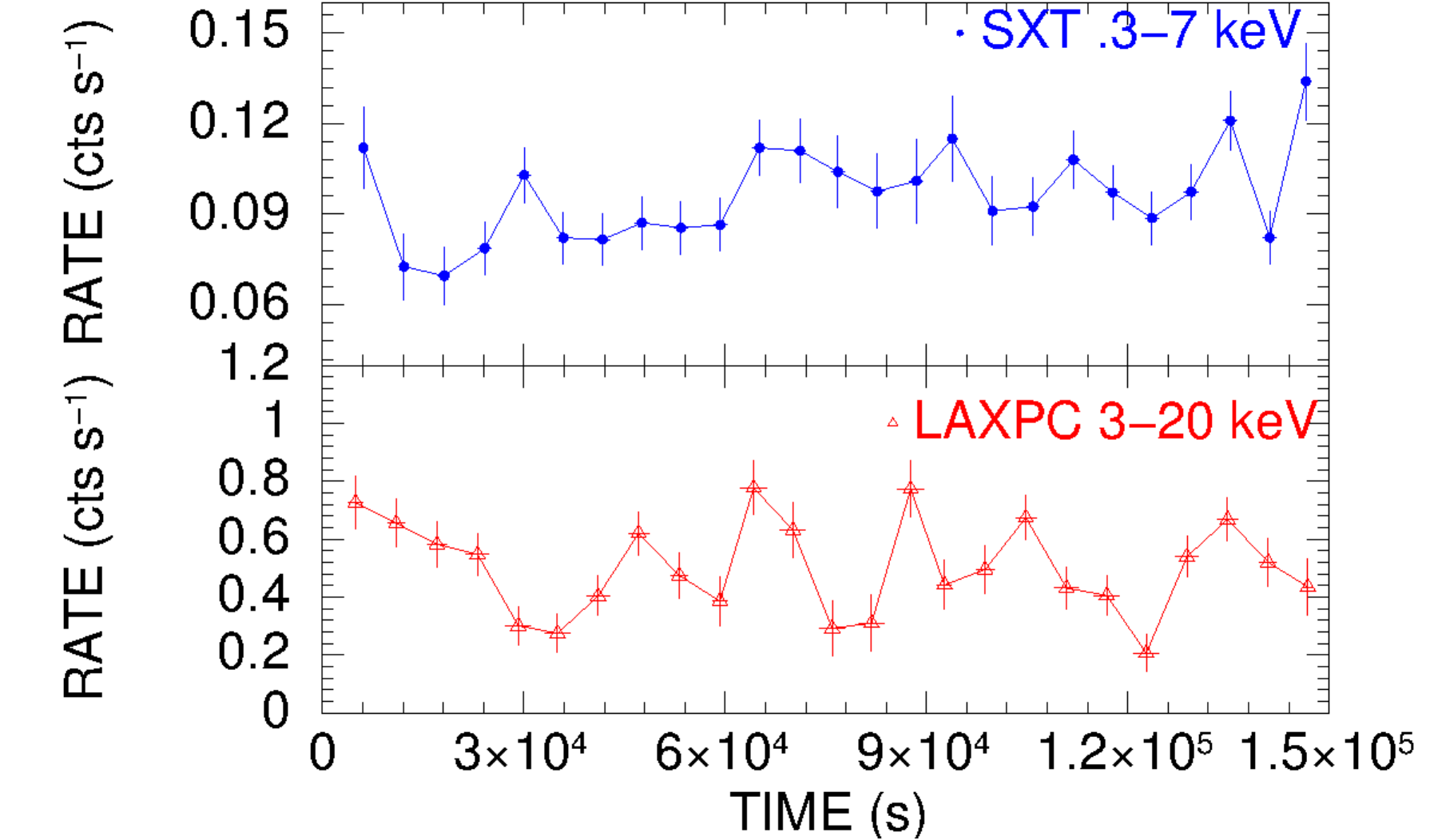}}
    \caption{Upper panel : Orbit-wise light curve of RE~J1034+396 in the energy range 0.3-7 keV from SXT observations. Lower panel : Orbit-wise light curve of RE~J1034+396 in the 3-20 keV band  from the top layer of LAXPC LX20.
    Time in seconds 
starts from  2016 April 21 03:00 (UTC) i.e. MJD 57499.125.  }
    \label{fig:olcsl20}
\end{figure}

Using  a total of 67.3 ks of useful data from the top layer of LX20 (LX20-L1) the light curve for the 3-20 keV band
shown in the lower panel of \Cref{fig:olcsl20} was generated.
For each orbit (with an average source exposure of $\sim 3$ ks) the normalised average background rate has been subtracted from the source count rate.
For this LAXPC time-series a fractional variability
amplitude of 0.278$\pm$0.053 is found.

We note that the data from each satellite orbit (shown in \Cref{fig:olcsl20}) is not strictly simultaneous
for the SXT and the LAXPC instruments, primarily because the  Earth Elevation angle data selection criterion is more stringent for the SXT.
The LAXPC light curve, however, has a significantly
higher fractional variability than that for the  SXT. 
%It is quite conceivable  that 
%the source shows higher variability at energies above 5 keV, though we cannot
%completely rule out the possibility of some residual background
%fluctuations in the LAXPC data.  

% Example figure

\subsection{Spectral Analysis}

X-ray spectra for the source and background regions for the entire SXT dataset were extracted
(as described in section 2.1).
Data from the LX20 top layer were used to generate orbit-wise
on-source and background spectra using the 
prescription given in section 2.2.  These were combined using appropriate scaling factors.
Further, these merged source and background spectra were compared in 35-80 keV region 
and normalised.

XSPEC compatible RMF and ARF files for the SXT and LAXPC were obtained from the ASTROSAT
Science Support Cell\footnote{\url{http://astrosat-ssc.iucaa.in/?q=data\_and\_analysis}}.
These were sxt\string_pc\string_excl01\string_v03.arf, sxt\string_pc\string_mat\string_g0to12.rmf
for the SXT and lx20cshm08L1v1.0.rmf for the LAXPC unit LX20.
SXT spectra were binned to a minimum of 20 counts per energy bin using
GRPPHA to facilitate $\chi^{2}$ fitting. XSPEC (version 12.9.0)
\citep{Arnaud_XSPEC_1996} was used. Energy ranges of 0.3-7 keV and 3-20 keV were used 
for the SXT and LX20 respectively. The SXT and LX20 data were fit simultaneously
with a  model including  components for the line of sight absorption, a soft excess and a power-law
tail similar to that used by \citet{Middle09}. Solar abundances were set to the most 
recent \emph{aspl} model \citep{aspl_2009}; photoelectric absorption cross-sections 
were set to \emph{vern}, and the default ${\rm \Lambda CDM}$ cosmology 
($<H_{0}>$=70,$<q_{0}>$=0.0,$<\Lambda_{0}>$=0.73) was used. Line of sight absorption
was modelled with Tuebingen-Boulder ISM absorption model (Tbabs) with the
$N_H$ value fixed at the galactic value of $1.47\times10^{20}$ cm$^{-2}$ \citep{middle11}.
The soft excess was modelled using the compTT model \citep{Titarchuk_comptt_1994}, for the 
Comptonization of soft photons in a hot plasma. A power-law was used to model 
continuum emission over the 0.3-20 keV range. To account for the relative normalisation
between the SXT and LX20, constant multiplicative factors for the two instruments were incorporated. 
{\bf The constant factor was fixed to 1 for the SXT and was allowed to vary and fitted for LX20.}
The auxiliary response file, 'sxt\_pc\_excl01\_v03.arf' used for the SXT 
excludes a circular region with a 1 arcmin radius centred on the source,
leading to an overestimation of the source flux. A correction factor for this effect has been estimated using SXT data from 
a bright source (1ES1959+650). It is found to be about 0.92 and this
correction is applied to the SXT flux estimate.

\begin{figure}
\centerline{\includegraphics[width=1.25 \linewidth]{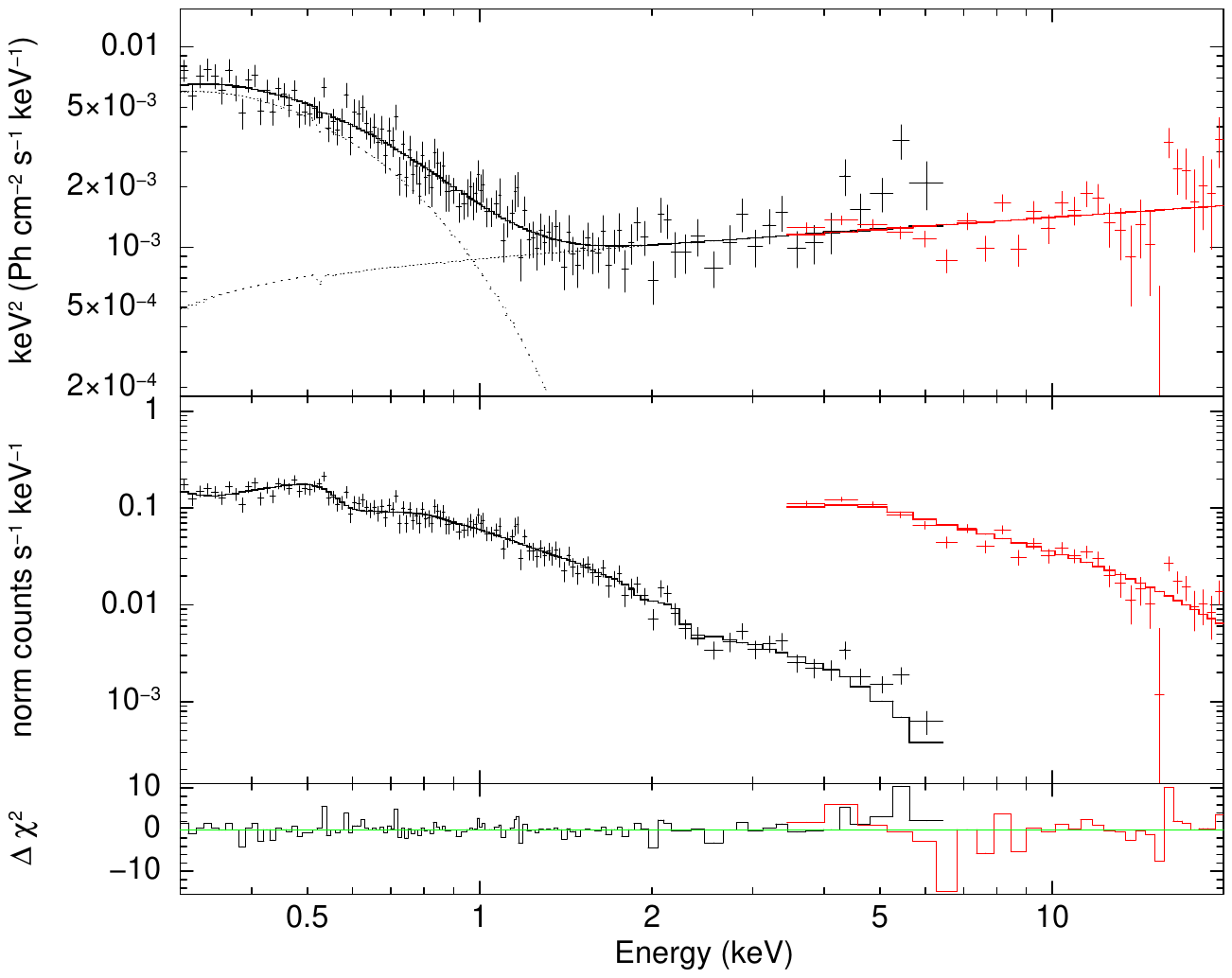}}
\caption{SXT and LAXPC LX20 layer 1 spectra of RE~J1034+396 fitted with the model constant*tbabs*(compTT+powerlaw).
The top panel shows the  spectral energy distribution $\nu f_{\nu}$, the middle panel shows the data and model fit while the bottom panel shows the residuals. The individual spectral components, CompTT and power-law, are separately shown in the top panel.}
\label{fig:astrosat-spec}
\end{figure}

A difference in the SXT gain function compared to that used in the response matrix revealed  by residuals near the gold absorption edge
at $\sim$ 2 keV was corrected by using the  gain fit command. While executing this command,
slope was fixed to 1 and offset was varied. {\bf A positive gain offset of 33 eV was seen} and
for the best fit a $\chi^{2}$/d.o.f.=212/146 was obtained {\bf with null hypothesis probability 
of 3.08 $\times$ 10$^{-4}$}. The best fit model
parameters and their 90\% confidence errors are given in Table~\ref{rej1034:tab:specfits}.
{\bf The relative normalisation between LX20 and the SXT  is $1.37_{-0.15}^{+0.17}$.}
The fitted spectra are shown in \Cref{fig:astrosat-spec} along with the spectral energy distribution
and residuals in terms of $\chi^{2}$.

\citet{Middle09} made a detailed spectral analysis of RE~J1034+396 using the XMM Newton data
when the high frequency QPO (HFQPO) was detected and concluded that the soft excess seen in this
source can be modelled as a  low-temperature Comptonization spectra with high optical depth, 
a feature also seen in black hole binaries  exhibiting super Eddington accretion like GRS~1915+105. \citet{middle11} presented an extensive spectral analysis of four XMM Newton observations with a similar spectral model. 
The measured spectral parameters presented in this work  are found to be consistent with those obtained by
\citet{middle11} for the third observation (see their Table 2),   though the power-law index is found to be  harder in our case (1.80 instead of 2.37).

\begin{table*}
  \begin{center}
 \caption{Best-fitting parameters for the model const*Tbabs*(compTT+po).
Error bars are at $90\%$ confidence, and ``peg'' indicates that the
parameter has reached its limit. $\Gamma$ is the photon
spectral index, $T_0$ is the seed soft photon (Wien) temperature in keV,
k$T_e$ is the plasma temperature in keV, $\tau$ is the plasma optical depth.  }
 \label{rej1034:tab:specfits}
 \begin{tabular}{lrrrrrrrrr}
 \hline
 \hline
\noalign{\smallskip}
 $\Gamma^a$              & $T_0$                & k$T_e$                    & $\tau$                 & const                  & $\chi^2$/dof & $F^{\mathrm{b}}$ \\
                         & (keV)                & (keV)                     &                        &                        &              & (1--10\,keV)   & \\
\noalign{\smallskip}
 \hline
\noalign{\smallskip}
$1.80 \pm 0.09$        & $0.03_{peg}^{+0.02}$ & $0.140_{-0.021}^{+0.025}$ & $21.157_{-4.40}^{peg}$ & $1.37_{-0.15}^{+0.17}$ & 212/146      & $4.05$  \\
\noalign{\smallskip}
  \hline
  \end{tabular}
  \end{center}
  \begin{list}{}{}
    \item[$^{\mathrm{a}}$]Photon index, $n_E \propto E^{-\Gamma}$ (ph cm$^{-2}$ s$^{-1}$\,keV$^{-1}$).
    \item[$^{\mathrm{b}}$]Flux in units of $10^{-12}$ erg cm$^{-2}$ s$^{-1}$.
  \end{list}
  \end{table*}

\section{Swift and XMM-Newton Observations}

To understand the source variability behaviour at different time scales, we have 
analysed the data from Swift (spanning timescales of days to months) and XMM-Newton
(covering the shorter time scales).  Here, for consistency, we only use low energy
data ($\sim$ 0.5 - 8 keV) for an analysis of variability and Power Spectral Density (PSD)
generation  at a variety of time scales.

\subsection{Swift   Observations}

\begin{figure}
\centerline{\includegraphics[width=0.72 \linewidth, angle=270]{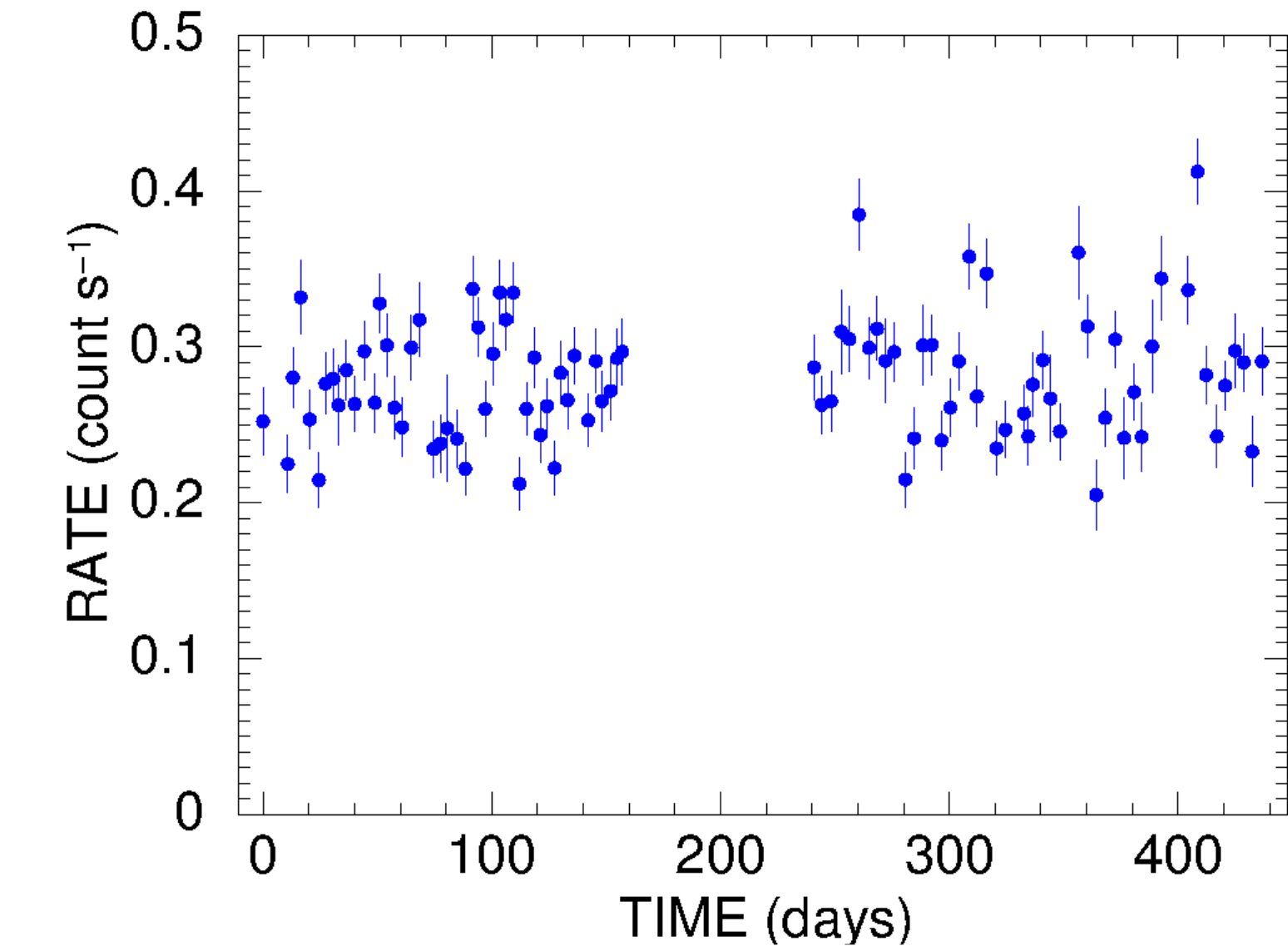}}
\caption{Light curve of RE~J1034+396  obtained from Swift-XRT observations     
during 2016-01-30 to 2017-04-10 in the 0.3 - 8 keV band.
Count rates are average values for each observation. 0 corresponds to MJD 57400.}
\label{fig:swxrtmjlc}
\end{figure}

RE~J1034+396 has been observed with the Swift XRT \citep{Burrows04} on several
occasions.  In the present work, we have used the data taken between   2016-01-30 
00:14:24 -- 2017-04-10 16:04:48, when the sampling was quite dense, obtained from the Swift public 
archive\footnote{\url{https://swift.gsfc.nasa.gov/archive/}}. All the data sets were
processed using the XRTDAS (version 3.3.0) software package available under HEASoft (version
6.21). The standard procedure involving xrtpipeline (version 0.13.3) was used for cleaning and 
calibrating event files. Data taken in Photon Counting (PC) mode were used  
and the standard grade selection of 1-12 was applied in the analysis. 
A circular region with a radius of 90 arcsec centered at the location of RE~J1034+396 was used 
to extract source counts.  A 180 arcsec circular  region 
offset  from the source region was used for background estimation. Source and background light
curves with bin sizes of 10 s and over the energy range of 0.3-8 keV were extracted using 
tool xrtproducts (version 0.4.2). The source light curves were corrected for telescope 
vignetting as well as for PSF losses using the tool {\it xrtlccorr} (version 0.3.8). Background subtracted
source light curves were then obtained. For each observation an average rate was calculated. 
The observation-wise light curve is shown in \Cref{fig:swxrtmjlc}.
Typically there is one observation $\sim $1 ks every 3-4 days.

 The fractional variability estimated for this light
curve is 0.119$\pm$0.012, very similar to that found with the SXT.

\subsection{XMM-Newton observation}

\begin{figure}
\centerline{\includegraphics[width=0.75 \linewidth, angle=270]{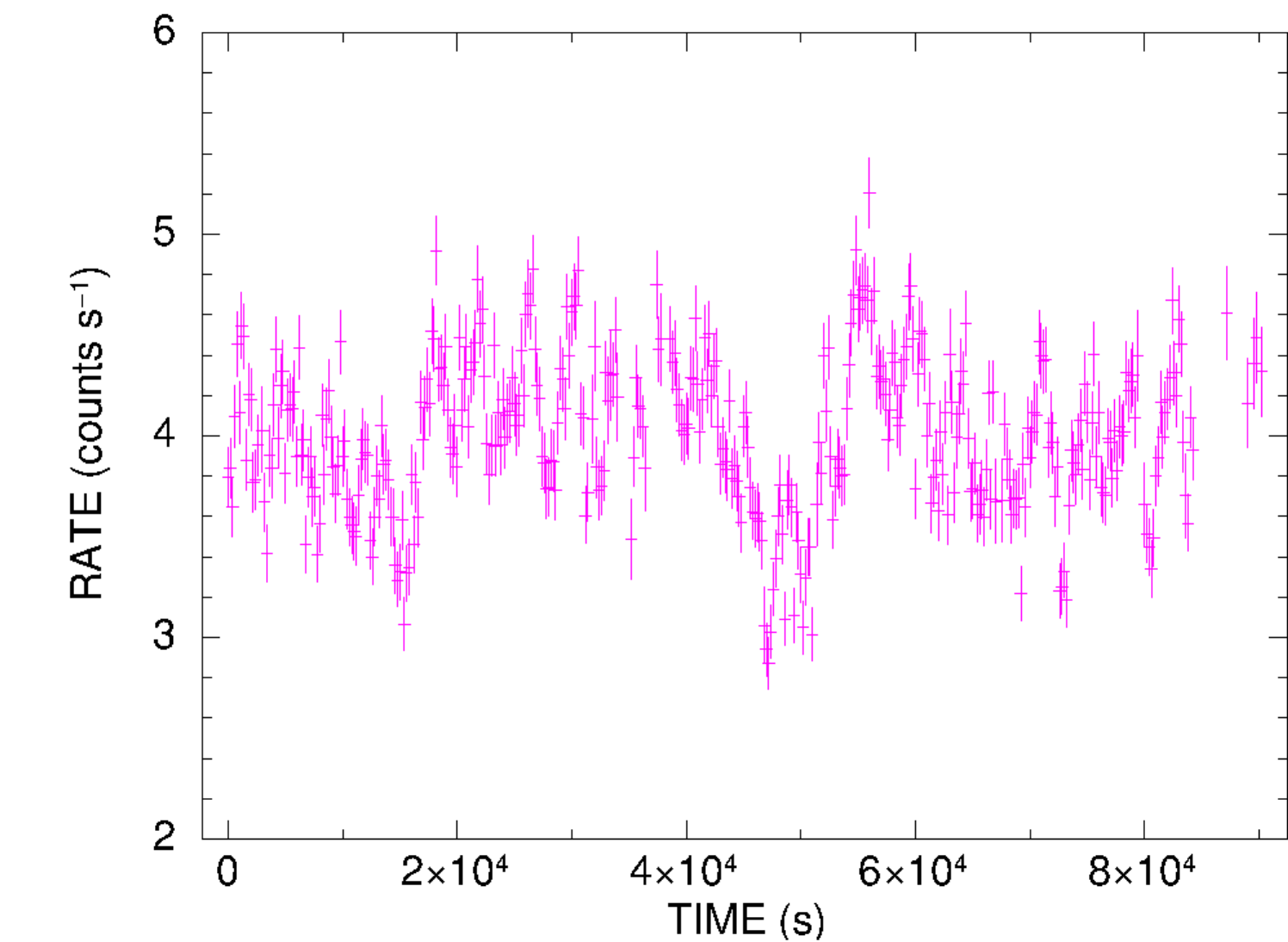}}
\caption{Light curve  of RE~J1034+396 from EPIC-PN on board XMM-Newton over the energy range 0.3-10 keV during 
observations carried out on 2007 May 31 starting at 20:10:12 (MJD 54251.840). Count rates are binned over 200 s.}
\label{fig:xmmlc}
\end{figure}

RE~J1034+396 was observed with XMM-Newton Satellite \citep{jansen01} on several occasions. We have
analysed the data from observations carried out on 2007 May 31, which is one of the longest data
stretches available with a duration of 93 ks \citep{Gierli08}. The XMM-Newton satellite has two 
X-ray instruments: (i) The European Photon Imaging Cameras (EPIC) and
(ii) The Reflection Grating Spectrometers (RGS).
There are three EPIC cameras, MOS1, MOS2 and PN.
We use only the data from the EPIC-PN \citep{struder01}. Data reduction was accomplished using the 
XMM-Newton Science Analysis System (SAS v:15.0.0) along with the recent calibration files.
Unflagged events (flag = 0) with PATTERN $\leqq$ 4 for the PN camera were used.  Intervals during  
soft proton flares were excluded by generating a GTI (Good Time Intervals) file above 10 keV for 
the full field with RATE $\leqq$ 0.85 counts $s^{-1}$, which gives the maximum signal-to-noise ratio.
The filtered event file was extracted using evselect (SAS tool) for the energy range 0.2-10 keV.
An image was extracted using XSELECT. Circular regions with 
a radius of 20 arcsec centred on the source and with a radius of 40 arcsec offset 
from the source were used to extract source and background counts, respectively.
%Source and background regions with radii of 20 and 40 arcsec respectively were used. 
Light curves for source and background regions 
over the energy range of 0.3-10 keV with a bin size of 200 s were generated. The background subtracted
light curve shown in \Cref{fig:xmmlc} has a fractional variability amplitude of  0.088$\pm$0.003.

The variability amplitudes measured on different time scales from the various instruments are shown in
Table 2. For the SXT and XMM-Newton data, we have also give the variability in two different
energy ranges: below 1 keV (where the soft excess is dominant) and above
1 keV (dominated by the power-law).  The AstroSat results for the variability are consistent with the strong energy dependence
pointed out by \citet{Middle09} and \citet{middle11}. The XMM-Newton 1-10 keV variability result which is intermediate to the low energy
(SXT, Swift, and XMM) results and the higher energy LAXPC band is also  consistent.

\begin{table*}
\begin{center}
\caption{Fractional Variability Amplitude (FVA) for RE~J1034+396 at different time scales obtained from different instruments}
\label{tab:var_tab}
\begin{tabular}{llcccr} % four columns, alignment for each
\hline
Instrument & Energy range (keV)& Observation dates & Duration & Bin Size & Variability\\ 
\hline
Swift-XRT  & 0.3 - 8.0 &  30 January 2016 - 10 April 2017 & 436 days    & 4 days & 0.119$\pm$0.012 \\
SXT   & 0.3 - 7.0      & 21 April 2016 & 42.7 ks  & ~ 97 minutes  & 0.123 $\pm$ 0.031 \\
    &  0.3 - 1.0 &  & & &   0.123$\pm$0.033 \\
  &  1.0  - 7.0 &  & & &    $<$0.433 \\  
LAXPC (LX20-L1) & 3 - 20 & 21 April 2016 & 67.3 ks & ~ 97 minutes & 0.278$\pm$0.053 \\
XMM-Newton & 0.3 - 10.0 & 31 May - 1 June 2007 & 93 ks  & 200 s & 0.088$\pm$0.003 \\  
&  0.3 - 1.0 &  & & &   0.084$\pm$0.003 \\
&  1.0 - 10.0 &  & & &   0.157$\pm$0.0084\\              
\hline
\end{tabular}
\end{center}
\end{table*}

\section{Discussion}

RE~J1034+396 is one of the rare AGN showing a clear X-ray QPO signal \citep{Gierli08}. The detection
of this QPO was facilitated by the fact that the observed QPO period ($\sim$1 hr) is quite
short. Comparing to GRS~1915+105, which has  a mass of $\sim 12.9\pm2.4$M$_\odot$ \citep{Hurley13},
%{\bf Hurley et al. (2013) (add in the reference list)}][]{},
this period could be the regular C-type QPO seen at a few Hz in the hard state of low mass black hole sources, 
\citep{Reig2000}, 
%{\bf (Reig et al. 2000 (add in the reference list))},
would imply  a black hole mass of $\sim$10$^5$M$_\odot$ in RE~J1034+396.
Alternatively, using the high frequency
QPO seen at $\sim 65-67$~Hz \citep{alta2012}, would imply a black hole mass of $\sim$10$^6$M$_\odot$.
%In the latter case, the mass of the black hole  in RE~J1034+396  would be  $\sim$7$\times$10$^6$M$_\odot$. 
One way to distinguish between these
two possibilities is to examine the power spectral density (PSD)  on a wider frequency range and compare it
with that seen from GRS~1915+105.
 The AstroSat data, combined with the other observations on diverse time scales, provides 
 this opportunity as shown below. 

We assume  that the general shape of PSD  does not change significantly
 on the time scale of a few years: a reasonable assumption considering the fact that typical spectral state changes in
 Galactic black hole sources occur on a time scale of several days, and scaling to supermassive 
 black holes of mass $\sim 10^5-10^6$~M$_\odot$, the spectral state change in RE~J1034+396 should occur in $\sim 100-1000$ years. 
 We, however, note here that even in the state where high frequency QPOs are produced, there could be subtle variations in the source behaviour such that the QPOs are seen only at certain ranges of hardness ratios \citep{alta2012}. Similar state dependent QPO observations have also been reported for RE~J1034+396 \citep{Alston14}.
 Even then, properties of the source such as  the  spectrum and PSD remain unchanged.

\begin{figure}
\centerline{\includegraphics[width=0.9 \linewidth]{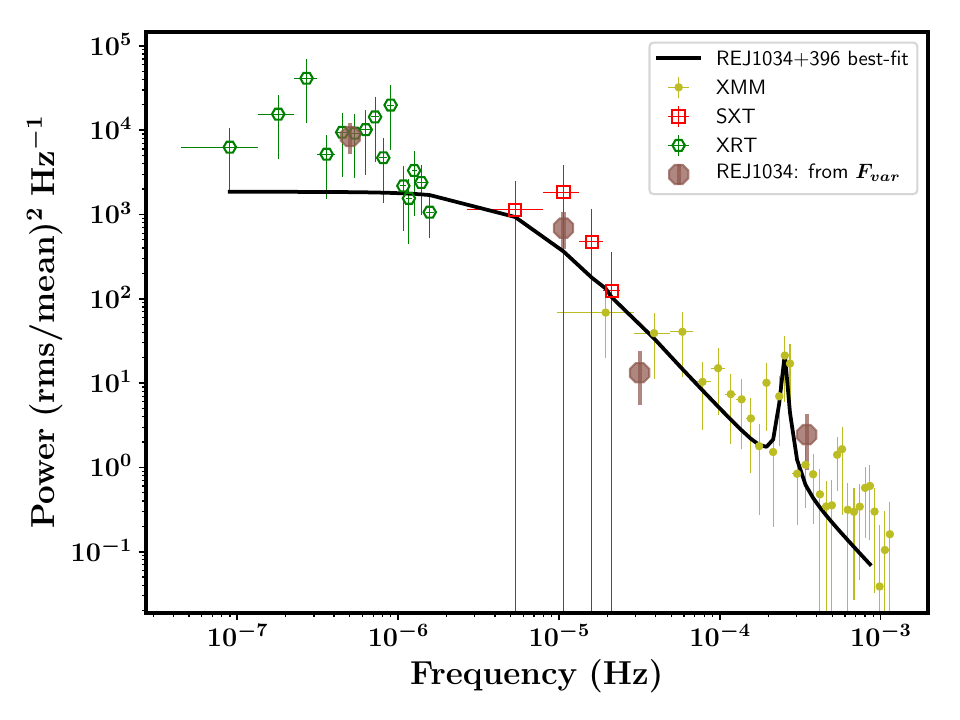}}
\caption{The power spectral density of RE~J1034+396 obtained using Swift XRT (green), AstroSat SXT (red) and XMM-Newton EPIC-PN (yellow) data.
	The brown points are calculated from the fractional variability amplitude $\rm{F_{var}}$ (Equation~\ref{Fvar}) for these three data sets, which show a clear increase of power at longer time scales. 
	The continuous black line is an best-fit model consisting of a zero-centered Lorentzian describing the flat-top noise and another Lorentzian for the narrow QPO component. 
\label{fig7}}
\end{figure}

 In Table 2, we have
 shown the variability amplitude at different time scales and  there is an indication of the
 variability increasing at longer time scales.
 To quantify this, we have generated the PSD for RE~J1034+396 
 (using the $powspec$ tool in $ftools$) and shown in Figure~\ref{fig7}. 
In this figure, the combined PSD using Swift XRT, AstroSat SXT and XMM-Newton EPIC-PN data is shown.
The fractional variability amplitude $\rm{F_{var}}$ (Equation~\ref{Fvar}) points for these three data sets are also
included, by converting them to the units of (rms/mean)$^2$ Hz$^{-1}$, by taking the average frequency from 
the bin size and data length (the Swift-XRT and XMM-Newton data are split into two bin sizes).  These points 
are consistent with the PSD points and clearly show an increase of power at longer time scales.
{\bf This type of behaviour was seen in several AGNs in the past, see for example, \citep{Uttley2002}.} 
{\bf We have fitted the PSD with a model consisting of a zero-centered Lorentzian to describe the
flat-top noise and another Lorentzian for the narrow QPO component. The reduced $\chi^2$ for this fit
is 0.81 for 41 degrees of freedom and the best fit QPO frequency is $2.53_{-0.08}^{+0.13}$ $\times$ 10$^{-4}$ Hz.
This is close to the value reported by \citet{Gierli08} and others.}
%Figure~\ref{fig7} also includes an eye-fit model consisting of a zero-centered Lorentzian to describe the 
%flat-top noise and another Lorentzian for the narrow $2.7\times10^{-4}$~Hz QPO component. 
Note that, while the flat-top noise description of the Swift XRT points appears to be reasonable, future
availability of PSD points in the frequency range of $\sim 10^{-6} - 10^{-5}$~Hz might be required
to verify this description.

\begin{figure}
\centerline{\includegraphics[width=0.9 \linewidth]{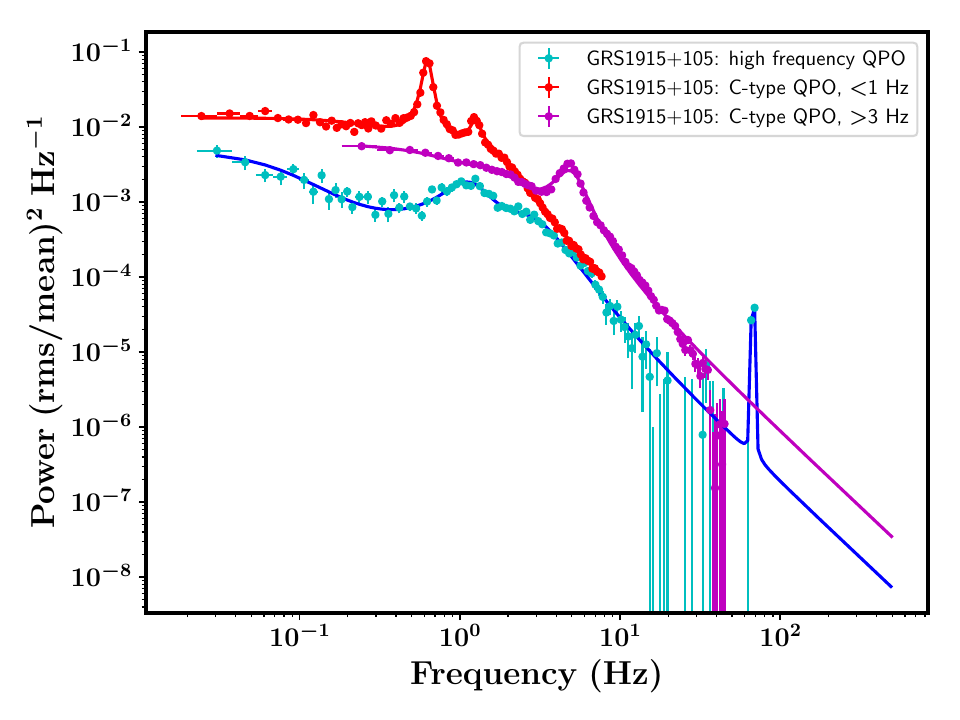}}
\caption{Power spectrum of GRS1915+105 for high frequency (cyan), C type lower frequency (red) and C type higher frequency
	(violet) QPOs respectively, along with the corresponding best-fit models \citep{Reig2000}. 
%{\bf (Reig et al. 2000 (add in the reference list), one reference for the high frequency QPO should be given)}.
\label{fig8}}
\end{figure}

For comparison, we have shown the PSDs of GRS 1915+105 in Figure~\ref{fig8}.
These PSDs are generated using data from the Proportional Counter Array (PCA) onboard the Rossi X-ray Timing Explorer 
(RXTE). GRS 1915+105 was observed on many occasions with RXTE. We have chosen the datasets corresponding
to high frequency QPO (observation ID: 80701-01-28-01 on 2003 October 21), C type QPO with lower frequency
(observation ID: 10408-01-27-00 on 1996 July 26) and C type QPO with higher frequency (observation ID:
10408-01-30-00 on  1996 August 18; \citep{Reig2000}).
%{\bf (Reig et al. 2000 (add in the reference list), one reference for the high frequency QPO should be given)}). 
The data were analysed using the standard procedure, light curves generated
using the tool SAEXTRCT under HEASoft version 6.21 and the PSD determined using the $powspec$ tool in $ftools$.

\begin{figure}
\centerline{\includegraphics[width=0.9 \linewidth]{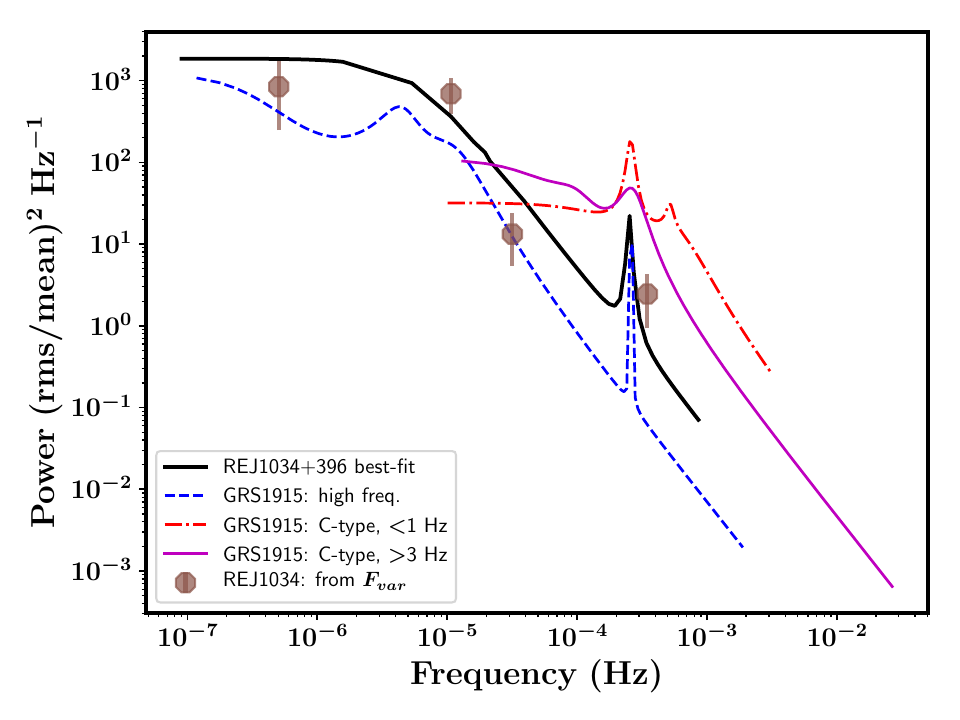}}
\caption{RE~J1034+396 power spectra (black line: best-fit model, brown points: spectrum calculated from $\rm{F_{var}}$; from Figure~\ref{fig7}),
	along with the different types of QPOs from GRS1915+105, scaled to match the QPO frequencies with that of RE~J1034+396. Blue, red
	and magenta lines correspond to rescaled power spectra for GRS1915+105 high frequency, C type lower frequency and C type higher
	frequency QPOs respectively (from Figure~\ref{fig8}). Note that, of the three, the power spectrum with the high frequency QPO from GRS1915+105 provides the best match with RE~J1034+396.
\label{fig9}}
\end{figure}

In Figure~\ref{fig9}, we compare the three GRS 1915+105 PSD best-fit model curves from Figure~\ref{fig8} with the
above-mentioned RE~J1034+396 {\bf best-fit} model curve from Figure~\ref{fig7} in the following way. The frequencies 
of each GRS 1915+105 PSD curve are scaled (divided) by a factor $k$, so that a GRS 1915+105 QPO appears at the RE~J1034+396
QPO frequency {\bf $2.53\times10^{-4}$~Hz}. Correspondingly, the powers of a GRS 1915+105 PSD are multiplied by the 
same factor.
Figure~\ref{fig9} shows that there is a marked similarity between the RE~J1034+396 PSD curve and the 
GRS 1915+105 PSD curve with the high frequency QPO, primarily in terms of the QPO width and the higher power at
the longer time scales. This is, however, not true for the GRS 1915+105 PSD curves for C type QPOs.
We, therefore, identify the RE~J1034+396 QPO as a high frequency QPO. The corresponding $k$-value is
{\bf $67.5/2.53\times10^{-4} = 266798$, implying a mass of $\sim (3.4\pm0.6)\times10^6$~M$_\odot$} for the black hole in RE~J1034+396.
{\bf This is consistent with the mass ranges for this object reported by \citet{Gierli08} for the
first time as well as by \citet{Czerny16} more recently.} 
%{\bf Czerny, B. et al. (2016) (add in the reference list)}.

   In conclusion, in this work we have presented the light curves and spectra of RE~J1034+396 obtained from the SXT and LAXPC
   instruments of AstroSat. We have found an indication of an increase in the variability as a function of energy and also at longer time scales. We have combined these results with other archival data and found a reasonable
   indication of a wide band PSD bearing a close resemblance to that seen in the Galactic black hole
   source, GRS 1915+105. We conclude that the QPO seen in RE~J1034+396 is more likely to be a high frequency QPO, and 
   use this to infer a  mass for  the black hole in RE~J1034+396.

%\begin{figure}
%\hskip 0.4in
%\begin{subfigure}[b]{0.37\textwidth}
%\centerline{\includegraphics[width=1.3 \linewidth]{rej1034_grs1915_composite.pdf}}
%\end{subfigure}

%\hskip 0.4in
%\begin{subfigure}[b]{0.37\textwidth}
%\centerline{\includegraphics[width=1.3 \linewidth]{grs1915_QPO.pdf}}
%\end{subfigure}
%\caption{Top panel : The power spectral density  of RE~J1034+396 obtained using  Swift XRT data (red),
%AstroSat SXT (green) and XMM-Newton EPIC-PN (magenta). The continuous black line is a model
%consisting of a zero centred Lorentzian describing the flat-top noise and 
%another Lorentzian for the QPO. Cyan, blue and violet correspond to rescaled PSDs for
%GRS1915+105 high frequency, C type lower frequency and C type higher frequency QPOs respectively.
%Bottom panel : Power spectrum of GRS1915+105 for high frequency (cyan), C type lower frequency
%(red) and C type higher frequency (violet) QPOs respectively.  }
%\label{fig:fig7a}
%\end{figure}

\section*{acknowledgements}

This publication uses  data from the AstroSat mission of the Indian Space Research Organisation (ISRO).
AstroSat observed this source as part of its SXT guaranteed time program for AGN studies (PI: K. P. Singh).
This work has used the data from the Soft X-ray Telescope (SXT) developed in a collaboration between TIFR, Mumbai,
and the University of Leicester and the Large X-ray Proportional Counter Array
(LAXPC) developed at TIFR, Mumbai. The SXT and LAXPC POCs at TIFR are thanked
for verifying and releasing the data via the ISSDC data archive and providing the necessary software tools.
This research has made use of the data and software obtained from NASAs High Energy Astrophysics Science Archive 
Research Center (HEASARC), a service of Goddard Space Flight Center and the Smithsonian Astrophysical 
Observatory. The use of the XRT Data Analysis Software (XRTDAS) developed under the responsibility of the
ASI Science Data Center (ASDC), is gratefully acknowledged. This work has made use of observations obtained 
with XMM-Newton, an ESA science mission with instruments and contributions directly funded by ESA member 
states and the USA (NASA). Partial support for the SXT UK-Indian collaboration was provided by the UKIERI programme
of the British council and Royal Society. 
Kishor Chaudhury expresses his profound gratitude to Dr. Samir Kumar Sarkar for teaching
the data analysis procedure and special thanks to Dr. Arunava Bhadra for his valuable suggestions.

\bibliographystyle{mnras}
\bibliography{rej1034_060418} % if your bibtex file is called example.bib

\begin{thebibliography}{}
\makeatletter
\relax
\def\mn@urlcharsother{\let\do\@makeother \do\$\do\&\do\#\do\^\do\_\do\%\do\~}
\def\mn@doi{\begingroup\mn@urlcharsother \@ifnextchar [ {\mn@doi@}
  {\mn@doi@[]}}
\def\mn@doi@[#1]#2{\def\@tempa{#1}\ifx\@tempa\@empty \href
  {http://dx.doi.org/#2} {doi:#2}\else \href {http://dx.doi.org/#2} {#1}\fi
  \endgroup}
\def\mn@eprint#1#2{\mn@eprint@#1:#2::\@nil}
\def\mn@eprint@arXiv#1{\href {http://arxiv.org/abs/#1} {{\tt arXiv:#1}}}
\def\mn@eprint@dblp#1{\href {http://dblp.uni-trier.de/rec/bibtex/#1.xml}
  {dblp:#1}}
\def\mn@eprint@#1:#2:#3:#4\@nil{\def\@tempa {#1}\def\@tempb {#2}\def\@tempc
  {#3}\ifx \@tempc \@empty \let \@tempc \@tempb \let \@tempb \@tempa \fi \ifx
  \@tempb \@empty \def\@tempb {arXiv}\fi \@ifundefined
  {mn@eprint@\@tempb}{\@tempb:\@tempc}{\expandafter \expandafter \csname
  mn@eprint@\@tempb\endcsname \expandafter{\@tempc}}}

\bibitem[\protect\citeauthoryear{{Agrawal} et~al.,}{{Agrawal}
  et~al.}{2017}]{Agrawal17}
{Agrawal} P.~C.,  et~al., 2017, \mn@doi [Journal of Astrophysics and Astronomy]
  {10.1007/s12036-017-9451-z}, \href
  {http://adsabs.harvard.edu/abs/2017JApA...38...30A} {38, 30}

\bibitem[\protect\citeauthoryear{{Alston}, {Markevi{\v c}i{\= u}t{\.e}},
  {Kara}, {Fabian}  \& {Middleton}}{{Alston} et~al.}{2014}]{Alston14}
{Alston} W.~N.,  {Markevi{\v c}i{\= u}t{\.e}} J.,  {Kara} E.,  {Fabian} A.~C.,
   {Middleton} M.,  2014, \mn@doi [\mnras] {10.1093/mnrasl/slu127}, \href
  {http://adsabs.harvard.edu/abs/2014MNRAS.445L..16A} {445, L16}

\bibitem[\protect\citeauthoryear{{Altamirano} \& {Belloni}}{{Altamirano} \&
  {Belloni}}{2012}]{alta2012}
{Altamirano} D.,  {Belloni} T.,  2012, \mn@doi [\apjl]
  {10.1088/2041-8205/747/1/L4}, \href
  {http://adsabs.harvard.edu/abs/2012ApJ...747L...4A} {747, L4}

\bibitem[\protect\citeauthoryear{{Antia} et~al.,}{{Antia}
  et~al.}{2017}]{Antia2017}
{Antia} H.~M.,  et~al., 2017, \mn@doi [\apjs] {10.3847/1538-4365/aa7a0e}, \href
  {http://adsabs.harvard.edu/abs/2017ApJS..231...10A} {231, 10}

\bibitem[\protect\citeauthoryear{{Arnaud}}{{Arnaud}}{1996}]{Arnaud_XSPEC_1996}
{Arnaud} K.~A.,  1996, in {Jacoby} G.~H.,  {Barnes} J.,  eds,  Astronomical
  Society of the Pacific Conference Series Vol. 101, Astronomical Data Analysis
  Software and Systems V. p.~17

\bibitem[\protect\citeauthoryear{{Asplund}, {Grevesse}, {Sauval}  \&
  {Scott}}{{Asplund} et~al.}{2009}]{aspl_2009}
{Asplund} M.,  {Grevesse} N.,  {Sauval} A.~J.,   {Scott} P.,  2009, \mn@doi
  [\araa] {10.1146/annurev.astro.46.060407.145222}, \href
  {http://adsabs.harvard.edu/abs/2009ARA%26A..47..481A} {47, 481}

\bibitem[\protect\citeauthoryear{{Bian} \& {Huang}}{{Bian} \&
  {Huang}}{2010}]{Bian10}
{Bian} W.-H.,  {Huang} K.,  2010, \mn@doi [\mnras]
  {10.1111/j.1365-2966.2009.15662.x}, \href
  {http://adsabs.harvard.edu/abs/2010MNRAS.401..507B} {401, 507}

\bibitem[\protect\citeauthoryear{{Breeveld} \& {Puchnarewicz}}{{Breeveld} \&
  {Puchnarewicz}}{1998}]{Breeveld98}
{Breeveld} A.~A.,  {Puchnarewicz} E.~M.,  1998, \mn@doi [\mnras]
  {10.1046/j.1365-8711.1998.01354.x}, \href
  {http://adsabs.harvard.edu/abs/1998MNRAS.295..568B} {295, 568}

\bibitem[\protect\citeauthoryear{{Burrows} et~al.,}{{Burrows}
  et~al.}{2004}]{Burrows04}
{Burrows} D.~N.,  et~al., 2004, in {Flanagan} K.~A.,  {Siegmund} O.~H.~W.,
  eds,  \procspie Vol. 5165, X-Ray and Gamma-Ray Instrumentation for Astronomy
  XIII. pp 201--216, \mn@doi{10.1117/12.504868}

\bibitem[\protect\citeauthoryear{{Chitnis}, {Pendharkar}, {Bose}, {Agrawal},
  {Rao}  \& {Misra}}{{Chitnis} et~al.}{2009}]{chitnis09}
{Chitnis} V.~R.,  {Pendharkar} J.~K.,  {Bose} D.,  {Agrawal} V.~K.,  {Rao}
  A.~R.,   {Misra} R.,  2009, \mn@doi [\apj] {10.1088/0004-637X/698/2/1207},
  \href {http://adsabs.harvard.edu/abs/2009ApJ...698.1207C} {698, 1207}

\bibitem[\protect\citeauthoryear{{Czerny} et~al.,}{{Czerny}
  et~al.}{2016}]{Czerny16}
{Czerny} B.,  et~al., 2016, \mn@doi [\aap] {10.1051/0004-6361/201628103}, \href
  {http://adsabs.harvard.edu/abs/2016A%26A...594A.102C} {594, A102}

\bibitem[\protect\citeauthoryear{{Done}, {Davis}, {Jin}, {Blaes}  \&
  {Ward}}{{Done} et~al.}{2012}]{Done12}
{Done} C.,  {Davis} S.~W.,  {Jin} C.,  {Blaes} O.,   {Ward} M.,  2012, \mn@doi
  [\mnras] {10.1111/j.1365-2966.2011.19779.x}, \href
  {http://adsabs.harvard.edu/abs/2012MNRAS.420.1848D} {420, 1848}

\bibitem[\protect\citeauthoryear{{Gierli{\'n}ski}, {Middleton}, {Ward}  \&
  {Done}}{{Gierli{\'n}ski} et~al.}{2008}]{Gierli08}
{Gierli{\'n}ski} M.,  {Middleton} M.,  {Ward} M.,   {Done} C.,  2008, \mn@doi
  [\nat] {10.1038/nature07277}, \href
  {http://adsabs.harvard.edu/abs/2008Natur.455..369G} {455, 369}

\bibitem[\protect\citeauthoryear{{Hurley}, {Callanan}, {Elebert}  \&
  {Reynolds}}{{Hurley} et~al.}{2013}]{Hurley13}
{Hurley} D.~J.,  {Callanan} P.~J.,  {Elebert} P.,   {Reynolds} M.~T.,  2013,
  \mn@doi [\mnras] {10.1093/mnras/stt001}, \href
  {http://adsabs.harvard.edu/abs/2013MNRAS.430.1832H} {430, 1832}

\bibitem[\protect\citeauthoryear{{Jansen} et~al.,}{{Jansen}
  et~al.}{2001}]{jansen01}
{Jansen} F.,  et~al., 2001, \mn@doi [\aap] {10.1051/0004-6361:20000036}, \href
  {http://adsabs.harvard.edu/abs/2001A%26A...365L...1J} {365, L1}

\bibitem[\protect\citeauthoryear{{Mason}, {Puchnarewicz}  \& {Jones}}{{Mason}
  et~al.}{1996}]{Mason96}
{Mason} K.~O.,  {Puchnarewicz} E.~M.,   {Jones} L.~R.,  1996, \mn@doi [\mnras]
  {10.1093/mnras/283.1.L26}, \href
  {http://adsabs.harvard.edu/abs/1996MNRAS.283L..26M} {283, L26}

\bibitem[\protect\citeauthoryear{{Middleton}, {Done}, {Ward}, {Gierli{\'n}ski}
  \& {Schurch}}{{Middleton} et~al.}{2009}]{Middle09}
{Middleton} M.,  {Done} C.,  {Ward} M.,  {Gierli{\'n}ski} M.,   {Schurch} N.,
  2009, \mn@doi [\mnras] {10.1111/j.1365-2966.2008.14255.x}, \href
  {http://adsabs.harvard.edu/abs/2009MNRAS.394..250M} {394, 250}

\bibitem[\protect\citeauthoryear{{Middleton}, {Uttley}  \& {Done}}{{Middleton}
  et~al.}{2011}]{middle11}
{Middleton} M.,  {Uttley} P.,   {Done} C.,  2011, \mn@doi [\mnras]
  {10.1111/j.1365-2966.2011.19185.x}, \href
  {http://adsabs.harvard.edu/abs/2011MNRAS.417..250M} {417, 250}

\bibitem[\protect\citeauthoryear{{Pfeffermann} et~al.,}{{Pfeffermann}
  et~al.}{1986}]{Pfeffer86}
{Pfeffermann} E.,  et~al., 1986, in Society of Photo-Optical Instrumentation
  Engineers (SPIE) Conference Series. pp 519--532, \mn@doi{10.1117/12.964956}

\bibitem[\protect\citeauthoryear{{Pounds} et~al.,}{{Pounds}
  et~al.}{1993}]{Pounds93}
{Pounds} K.~A.,  et~al., 1993, \mn@doi [\mnras] {10.1093/mnras/260.1.77}, \href
  {http://adsabs.harvard.edu/abs/1993MNRAS.260...77P} {260, 77}

\bibitem[\protect\citeauthoryear{{Pounds}, {Done}  \& {Osborne}}{{Pounds}
  et~al.}{1995}]{Pounds95}
{Pounds} K.~A.,  {Done} C.,   {Osborne} J.~P.,  1995, \mn@doi [\mnras]
  {10.1093/mnras/277.1.L5}, \href
  {http://adsabs.harvard.edu/abs/1995MNRAS.277L...5P} {277, L5}

\bibitem[\protect\citeauthoryear{{Puchnarewicz}, {Mason}, {Siemiginowska}  \&
  {Pounds}}{{Puchnarewicz} et~al.}{1995}]{Puchnar95}
{Puchnarewicz} E.~M.,  {Mason} K.~O.,  {Siemiginowska} A.,   {Pounds} K.~A.,
  1995, \mn@doi [\mnras] {10.1093/mnras/276.1.20}, \href
  {http://adsabs.harvard.edu/abs/1995MNRAS.276...20P} {276, 20}

\bibitem[\protect\citeauthoryear{{Puchnarewicz} et~al.,}{{Puchnarewicz}
  et~al.}{1997}]{Puchnar97}
{Puchnarewicz} E.~M.,  et~al., 1997, \mn@doi [\mnras]
  {10.1093/mnras/291.1.177}, \href
  {http://adsabs.harvard.edu/abs/1997MNRAS.291..177P} {291, 177}

\bibitem[\protect\citeauthoryear{{Puchnarewicz}, {Mason}  \&
  {Siemiginowska}}{{Puchnarewicz} et~al.}{1998}]{Puchnar98}
{Puchnarewicz} E.~M.,  {Mason} K.~O.,   {Siemiginowska} A.,  1998, \mn@doi
  [\mnras] {10.1046/j.1365-8711.1998.01286.x}, \href
  {http://adsabs.harvard.edu/abs/1998MNRAS.293L..52P} {293, L52}

\bibitem[\protect\citeauthoryear{{Puchnarewicz}, {Mason}, {Siemiginowska},
  {Fruscione}, {Comastri}, {Fiore}  \& {Cagnoni}}{{Puchnarewicz}
  et~al.}{2001}]{Puchnar01}
{Puchnarewicz} E.~M.,  {Mason} K.~O.,  {Siemiginowska} A.,  {Fruscione} A.,
  {Comastri} A.,  {Fiore} F.,   {Cagnoni} I.,  2001, \mn@doi [\apj]
  {10.1086/319775}, \href {http://adsabs.harvard.edu/abs/2001ApJ...550..644P}
  {550, 644}

\bibitem[\protect\citeauthoryear{{Reig}, {Belloni}, {van der Klis},
  {M{\'e}ndez}, {Kylafis}  \& {Ford}}{{Reig} et~al.}{2000}]{Reig2000}
{Reig} P.,  {Belloni} T.,  {van der Klis} M.,  {M{\'e}ndez} M.,  {Kylafis}
  N.~D.,   {Ford} E.~C.,  2000, \mn@doi [\apj] {10.1086/309469}, \href
  {http://adsabs.harvard.edu/abs/2000ApJ...541..883R} {541, 883}

\bibitem[\protect\citeauthoryear{{Serlemitsos} et~al.,}{{Serlemitsos}
  et~al.}{1995}]{Serlem95}
{Serlemitsos} P.~J.,  et~al., 1995, \pasj, \href
  {http://adsabs.harvard.edu/abs/1995PASJ...47..105S} {47, 105}

\bibitem[\protect\citeauthoryear{{Singh} et~al.,}{{Singh}
  et~al.}{2014}]{Singh14}
{Singh} K.~P.,  et~al., 2014, in Space Telescopes and Instrumentation 2014:
  Ultraviolet to Gamma Ray. p. 91441S, \mn@doi{10.1117/12.2062667}

\bibitem[\protect\citeauthoryear{{Singh} et~al.,}{{Singh} et~al.}{2016}]{kp16}
{Singh} K.~P.,  et~al., 2016, in Space Telescopes and Instrumentation 2016:
  Ultraviolet to Gamma Ray. p. 99051E, \mn@doi{10.1117/12.2235309}

\bibitem[\protect\citeauthoryear{{Singh} et~al.,}{{Singh} et~al.}{2017}]{kp17}
{Singh} K.~P.,  et~al., 2017, \mn@doi [Journal of Astrophysics and Astronomy]
  {10.1007/s12036-017-9448-7}, \href
  {http://adsabs.harvard.edu/abs/2017JApA...38...29S} {38, 29}

\bibitem[\protect\citeauthoryear{{Str{\"u}der} et~al.,}{{Str{\"u}der}
  et~al.}{2001}]{struder01}
{Str{\"u}der} L.,  et~al., 2001, \mn@doi [\aap] {10.1051/0004-6361:20000066},
  \href {http://adsabs.harvard.edu/abs/2001A%26A...365L..18S} {365, L18}

\bibitem[\protect\citeauthoryear{{Titarchuk}}{{Titarchuk}}{1994}]{Titarchuk_comptt_1994}
{Titarchuk} L.,  1994, \mn@doi [\apj] {10.1086/174760}, \href
  {http://adsabs.harvard.edu/abs/1994ApJ...434..570T} {434, 570}

\bibitem[\protect\citeauthoryear{{Uttley}, {McHardy}  \& {Papadakis}}{{Uttley}
  et~al.}{2002}]{Uttley2002}
{Uttley} P.,  {McHardy} I.~M.,   {Papadakis} I.~E.,  2002, \mn@doi [\mnras]
  {10.1046/j.1365-8711.2002.05298.x}, \href
  {http://adsabs.harvard.edu/abs/2002MNRAS.332..231U} {332, 231}

\bibitem[\protect\citeauthoryear{{Vaughan}, {Edelson}, {Warwick}  \&
  {Uttley}}{{Vaughan} et~al.}{2003}]{vaughan03}
{Vaughan} S.,  {Edelson} R.,  {Warwick} R.~S.,   {Uttley} P.,  2003, \mn@doi
  [\mnras] {10.1046/j.1365-2966.2003.07042.x}, \href
  {http://adsabs.harvard.edu/abs/2003MNRAS.345.1271V} {345, 1271}

\bibitem[\protect\citeauthoryear{{Yadav} et~al.,}{{Yadav}
  et~al.}{2016}]{Yadav16}
{Yadav} J.~S.,  et~al., 2016, in Space Telescopes and Instrumentation 2016:
  Ultraviolet to Gamma Ray. p. 99051D, \mn@doi{10.1117/12.2231857}

\makeatother
\end{thebibliography}

% Don't change these lines
\bsp    % typesetting comment
\label{lastpage}

\end{document}